# Thixotropy, non-monotonic stress relaxation, and the second law of thermodynamics

Yogesh M. Joshi

Department of Chemical Engineering, Indian Institute of Technology Kanpur, INDIA.

joshi@iitk.ac.in

**Abstract:**

Many thixo-viscoelastic materials have been reported to undergo enhancement in elastic modulus with time and decrease in the same under application of deformation field. Incorporation of this feature in a viscoelastic structural kinetic model has an apparent possibility of violating the second law of thermodynamics. Furthermore, in a related experimental observation, stress has been reported to undergo a non-monotonic change as a function of time under the application of constant strain. We analyze both these scenarios through a structural kinetic model that shows viscoelastic aging but undergoes rejuvenation only due to the viscous rate of strain. We observe that such formalism does not violate the second law. Interestingly, the proposed simple formalism predicts the experimental observation of the non-monotonic stress relaxation very well.

## I. Introduction

Thixotropy concerns the microstructural evolution of a soft material under quiescent or weak flow conditions causing a time-dependent increase in the viscosity and destruction of the evolved microstructure under strong flow conditions leading to a time-dependent decrease in the viscosity [1-5]. In a thixotropic material, spontaneous microstructural buildup, even under stress-free conditions, not just causes an increase in the viscosity but also that of elastic modulus. On the other hand, the application of the deformation field causes a decrease in the modulus. Furthermore, if such a material remains under quiescent conditions for a long time, the material may show yield stress causing the viscosity to tend to infinity [6-9], while the elastic modulus may continue to increase with time. An apparent possibility of the violation of the second law of thermodynamics by incorporation of such features in a structural kinetic model has been debated in the literature [10, 11]. In this work, we analyze this problem using a viscoelastic structural kinetic model. We also study its implications on experimentally observed non-monotonic stress relaxation recently reported in the literature [12].

The thermodynamically out of equilibrium soft materials, wherein constituents of the same get structurally arrested in physical or energetic cages formed by their neighbors, render them limited access to the phase space [13]. Any out of equilibrium material, by virtue of thermal motions of its constituents, undergoes microstructural evolution as a function of time and lowers the free energy. This process has been represented as physical aging in contemporary literature, and this class of materials has been termed soft glassy materials [14]. Physical aging causes the stress relaxation time of the same to increase as a function of time [15, 16]. Furthermore, depending upon the microstructure of a soft glassy material, the elastic modulus may also increase with time. However, when the elastic modulus shows an increase with time, the rate of growth has been observed to be much weaker than the corresponding increase in relaxation time [17]. Depending on the intensity of increase in relaxation time on time, material may eventually show yield stress when kept under quiescent conditions, and the corresponding viscosity approaches infinity [7]. However, the elastic modulus continues to evolve indefinitely. Such a state of a material can be considered as a limiting state of thixo-viscoelastic material with a dominant elastic component. The structural breakdown in the soft glassy material under application of deformation field has been termed rejuvenation, which, in principle, reverses the structure formed during physical aging [18, 19]. Consequently, the application of the deformation field causes a time-dependent decrease in viscosity, elastic modulus, and relaxation time. However, more complex situations have been reported in the literature, wherein application of moderate-intensity of deformation field alters the relaxation time distribution in an intricate fashion, causing a time-dependent increase in mean relaxation time. This phenomenon has been represented as overaging [20, 21].

In a seminal review on modeling thixotropic materials, Larson [10] considered a structural kinetic formalism along with the Hooke's law to propose a constitutive equation of thixoelastic material. He proposed a kinetic equation for change in structure parameter $\Lambda$, given by:

$$\frac{d\Lambda}{dt^*} = k_+(1-\Lambda) - k_-\Lambda\dot{\gamma}^*, \tag{1}$$

where $k_+$ and $k_-$ are the parameters associated with, respectively, the structural growth (physical aging) and breakdown (rejuvenation) terms, $t^*$ is the time and $\dot{\gamma}^*$ is the second invariant of the rate of the strain tensor. The structure parameter $\Lambda$, which represents the extent of microstructural evolution, varies between two limits: 0 (the state devoid of structure) and 1 (the state with a fully developed structure). Eq. (1) is combined with Hooke's law,

$$\sigma^* = G(\Lambda)\gamma, \tag{2}$$

where $\sigma^*$ and $\gamma$ are respectively the second invariant of stress and strain tensor. In Eq. (2) thixotropic modulus $G(\Lambda)$ always depends monotonically on $\Lambda$. Larson [10] assumed this relationship to be $G(\Lambda) = G_0\Lambda$. In order to verify thermodynamic consistency, it is necessary to assess that the above formalism does not violate a relevant statement of the second law. For the present scenario, the most pertinent statement of the second law of thermodynamics is due to Planck, and is given by "*It is impossible to construct an engine which will work in a complete cycle, and produce no effect except the raising of a weight and the cooling of a heat-reservoir*" [22]. Larson [10] considered a thought experiment, wherein material represented by Eq. (1) and (2), in a state with $\Lambda \approx 1$ is subjected to fast flow such that $\Lambda$ reduces rapidly below unity as the structure undergoes breakdown according to Eq. (1). At this state, the material is maintained under quiescent conditions ($\dot{\gamma}^*=0$), so that $\Lambda$ undergoes evolution following Eq. (1) and eventually attains a high value of $\Lambda$ that is close to unity (the initial state). Subsequently, the flow is entirely reversed by applying a specific magnitude of strain rate such that $\Lambda$ remains at the same value, thereby completing a cycle. Larson [10] argued that during the forward path due to structural breakdown, a decrease in $\Lambda$ would lead to a smaller value of $G(\Lambda)$ while during the reverse direction, since $\Lambda$ is high, $G(\Lambda)$ will have a high value. Consequently, the material will perform more work on the surrounding in the reverse path than what was done on it during the forward path, leading to the net effect of raising of a weight at the end of a cycle. Since this violates the second law of thermodynamics, Larson [10] argued that ideal thixoelasticity might not exist.

In the literature, different kinds of structural kinetic models have been proposed. The variety of system-specific as well as generic expressions of structural buildup and breakdown terms used in the kinetic expressions of the structure parameter, which is equivalent to Eq. (1), have been discussed in the review papers and monographs [1, 2, 5, 10]. Under no-flow conditions, these models predict the spontaneous evolution of the structure parameter. Such spontaneous change suggests the out - of - equilibrium nature of the fluid. Considering the fluid to be under constant temperature ($T$) and constant pressure ($P$) conditions, spontaneous evolution of the same indicates lowering of the Gibbs free energy. On the other hand, the application of the deformation field causes a decrease in the structure parameter due to progressive breakdown of the structure, thereby increasing the Gibbs free energy. This equivalence between increase (decrease) in structure parameter and decrease (increase) in free energy implies monotonically decreasing dependence of the Gibbs free energy on the structure parameter.

The majority of models employ two kinds of constitutive relations, either generalized Newtonian fluid that characterizes inelastic behavior or linear Maxwell model that represents viscoelastic behavior. In the Maxwell model, a limiting case of viscosity tending to infinity leads to Hooke's law represented by Eq. (2). This suggests that not just the so-called thixoelastic materials but also thixoviscoelastic materials, for a particular parameter range, could also be prone to violation of the second law as per Larson's argument [10]. Consequently, if Eq. (2) is replaced by the Maxwell model, then it is important to understand the threshold of viscosity beyond which the second law is expected to get violated.

Another important aspect that needs evaluation is the nature of the rejuvenation term in Eq. (1), wherein the structure parameter for a purely elastic material depends on the rate of strain. Since the decrease in $\Lambda$ suggests structural breakdown that implies dissipation, whether such behavior is possible for a purely elastic material needs to be analyzed. There are few structural kinetic models in the literature, wherein rejuvenation term has been considered to be dependent on the dissipative part (either viscous or plastic) of the rate of strain [8, 23, 24]. In a structural kinetic model based on fluidity (increase in fluidity can be considered as a decrease in structure parameter), Joshi [7] argued that rejuvenation is necessarily caused only by the strain rate associated with dissipative processes. It would, therefore, be instructive to analyze the consideration of viscous strain rate in the structural breakdown (rejuvenation) term of the kinetic equation on the violation of the second law of thermodynamics.

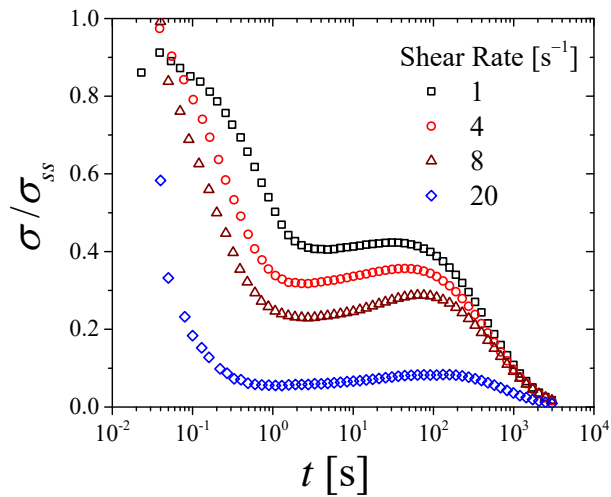

**Figure 1.** Stress relaxation (after the stoppage of steady-state shear flow) under constant strain conditions is plotted as a function of time for an aqueous solution of two polymers. Stress is normalized by shear stress at the steady-state ($\sigma_{SS}$). The legend represents four steady-state shear rates before the stress relaxation. The experimental data is taken from Fig. 2(c) of Hendricks and coworkers [12].

Recently, Hendricks and coworkers [12] studied three different gel-forming systems and observed that stress relaxation subsequent to cessation of steady shear flow follows a non-monotonic dependence on time. As shown in Fig. 1, in a typical observation, stress is observed to decay on the expected lines immediately after stopping the shear flow. However, over some intermediate timescales, stress is observed to show a noticeable increase before eventually decreasing to a completely relaxed state. The authors also propose a phenomenological model based on the redistribution of energy upon cessation of flow. They propose that the bond formation in partially aligned elastic domains during the structural reformation leads to a local increase in elastic energy that results in the observed rise in stress. Interestingly Joshi [7] was the first to suggest the possibility of such non-monotonic stress relaxation through a fluidity model, which he attributed to enhancement of elastic modulus during structural recovery (physical aging). While Hendricks and coworkers [12] present a convincing argument on how such nonmonotonicity does not violate the laws of thermodynamics, we feel it is important to analyze this behavior from a structural kinetic formalism.

In this work, we consider a structural kinetic model that is viscoelastic in nature. We independently consider two types of rejuvenation contributions, total strain rate, and viscous strain rate. We then test the validity of the second law of thermodynamics by applying series of deformation fields to the models. We also explore the possibility of non-monotonic stress relaxation subsequent to steady shear using the proposed formalism.

**II. Structural Kinetic Model**

The structural kinetic models are the most straightforward mathematical representations of the thixotropic behavior. Typically, structural kinetic models have three ingredients: a kinetic equation of the structure parameter, a constitutive relation, and a relationship between the parameters of the constitutive equations (viscosity, modulus, etc.) and the structure parameter [25]. In this work, we use the same expression for the evolution of structure parameter $\lambda$ that represents the extent to which the structure is formed during microstructural evolution or physical aging as proposed by Coussot and coworkers [6, 26] and is given by:

$$\frac{d\lambda}{dt^*} = \frac{1}{T_0} - \beta\lambda\dot{\gamma}^*, \tag{3}$$

where the first term on the right suggests structure formation with $T_0$ being the thixotropic timescale, the second term represents structure breakdown, wherein $\beta$ is a parameter. In this paper, as a convention, we describe the dimensional variables with superscript $*$. Eq. (3) considers that rejuvenation or the break-down of the structure evolved during physical aging occurs due to total rate of deformation induced in a material ($\dot{\gamma}^*$). A part of the total rate of strain is a rate of change of elastic strain. The corresponding elastic strain induced in material, however, can be completely recovered upon removal of stress. Consequently, in an independent proposal, we assume that the rejuvenation or the microstructural break-down due to the applied deformation field cannot be associated with the elastic strain rate. Therefore, we propose an alternate form of kinetic expression wherein the rejuvenation term in Eq. (3) is dependent only on the viscous strain rate ($\dot{\gamma}_v^*$, or the second invariant of the rate of strain tensor associated with the viscous flow) and is given by:

$$\frac{d\lambda}{dt^*} = \frac{1}{T_0} - \beta\lambda\dot{\gamma}_v^*, \tag{4}$$

In both Eqs. (3) and (4), $\lambda$ varies from 0 (devoid of structure) to $\infty$ (fully developed structure). In addition, we use the time-dependent linear Maxwell model as a constitutive equation, given by:

$$\dot{\underline{\gamma}}^* = \dot{\underline{\gamma}}_v^* + \dot{\underline{\gamma}}_e^* = \frac{\underline{\sigma}^*}{\eta(\lambda)} + \frac{d}{dt^*}\left(\frac{\underline{\sigma}^*}{G(\lambda)}\right), \tag{5}$$

where $\dot{\underline{\gamma}}_v^*$ and $\dot{\underline{\gamma}}_e^*$ are the rate of strain tensors respectively associated with the elastic and viscous elements. Furthermore, $\eta(\lambda)$ and $G(\lambda)$ are, respectively, the viscosity of dashpot and the modulus of spring associated with the Maxwell model. Both viscosity and modulus depend on time through their dependence on structure parameter $\lambda$. The relaxation time ($\tau$) of the Maxwell model is related to $\eta(\lambda)$ and $G(\lambda)$ as: $\eta(\lambda) = \tau(\lambda)G(\lambda)$. In absence of rejuvenation, both Eqs. (3) as well as (4), lead to, $\lambda \sim t^*/T_0$. On the other hand, under no-flow conditions, the relaxation time of several diverse types of soft glassy materials has been observed to show a power-law dependence on waiting time given by: $\tau \sim t^{*\mu}$, where $\mu$ is the power-law exponent [13, 17, 18, 27]. Considering this, we propose the dependence of $\tau$ on $\lambda$ given by:

$$\tau = \tau_0 \tilde{\tau}(\lambda) = \tau_0 \left(1 + \lambda^\mu\right), \tag{6}$$

where $\tau_0$ is the relaxation time associated with a completely rejuvenated state ($\lambda = 0$).

In a soft glassy material, wherein elements constituting the same are trapped in energy well having depth $U$, the relaxation time of the same is considered to show Arrhenius dependence on $U$ as [28]: $\tau = \tau' \exp(U/kT)$. Suppose in a completely rejuvenated state, if an energy well depth of an element is $U_0$, the relationship between $\tau_0$ and $\tau'$ is given by: $\tau_0 = \tau' \exp(U_0/kT)$. On the other hand, if $b$ is the characteristic length-scale associated with a soft glassy material, the elastic modulus of the same is related to $U$ as [28]: $G \approx U/b^3$. Both these relationships lead to a natural dependence of modulus on the relaxation time given by:

$$G = G_0 \tilde{G}(\lambda) = G_0 \left[1 + g \ln(\tilde{\tau})\right] = G_0 \left[1 + g \ln(1 + \lambda^\mu)\right], \tag{7}$$

where $G_0$ is the modulus associated with a completely rejuvenated state. The parameter $g = 1/\ln(\tau_0/\tau')$ controls the growth of modulus as a function of $\lambda$. Since in a completely rejuvenated or shear melted state, a material is devoid of any structure, we consider the barrier height $U_0$ to be typically of the order of thermal energy $U_0/kT = O(1)$, leading to $g = 1$.

In the present work, we solve two types of structural kinetic models. In the first formalism, which is termed as total strain rate (TSR) model, we use Eqs (5) to (7) along with Eq. (3). In the second formalism, which we call viscous strain rate (VSR) model, we use Eqs (5) to (7) along with Eq. (4). We employ the following non-dimensionalization scheme: $\dot{\gamma} = T_0 \dot{\gamma}^*$, $t = t^*/T_0$ and $\underset{\sim}{\sigma} = \underset{\sim}{\sigma}^*/G_0$. The conventional expression for the evolution of $\lambda$, given by Eq. (3) that we use in TSR model can be expressed in a dimensionless form as:

$$\frac{d\lambda}{dt} = 1 - \beta \lambda \dot{\gamma}. \tag{8}$$

On the other hand, the newly proposed expression for the evolution of $\lambda$ associated with VSR model, wherein rejuvenation depends on the viscous strain rate is given by Eq. (4). In a dimensionless form Eq. (4) is represented as:

$$\frac{d\lambda}{dt} = 1 - \beta\lambda\dot{\gamma}_v = 1 - \alpha\frac{\beta\lambda\sigma}{\tilde{\tau}(\lambda)\tilde{G}(\lambda)}, \tag{9}$$

where $\sigma$ is the second invariant of the stress tensor. Finally, the time dependent Maxwell model given by Eq. (5) in a dimensionless form can be expressed as:

$$\dot{\underline{\gamma}} = \dot{\underline{\gamma}}_v + \dot{\underline{\gamma}}_e = \alpha\frac{\underline{\sigma}}{\tilde{\tau}(\lambda)\tilde{G}(\lambda)} + \frac{d}{dt}\left[\frac{\underline{\sigma}}{\tilde{G}(\lambda)}\right]. \tag{10}$$

In Eqs. (9) and (10), $\alpha = T_0/\tau_0$ is the dimensionless variable describing the ratio of the thixotropic timescale to the relaxation time associated with the completely rejuvenated state. Consequently, $\alpha \to 0$ is the purely elastic limit of the Maxwell model. The model has four parameters: $\beta$, $g$, $\mu$ and $\alpha$. In this work, we always consider $\beta = 1$ and $g = 1$. We also consider $\mu = 1.2$ unless otherwise specified. Furthermore, in this work, we consider only the shear flow, wherein the velocity is in direction 1, and the gradient is in the orthogonal direction 2.

In the present paper, we analyze the energy required to deform the structural kinetic model under the application of constant strain rate and compare the same with the work that can be recovered from it when kept under constant strain conditions as a function of time. This flow field allows us to assess the models with respect to Planck's statement of the second law. We also investigate the evolution of the stress as a function of time under constant strain conditions.

### III. Results and discussion

We first discuss the steady-state behavior of the model. In a limit of the steady-state, the time-dependent Maxwell model represented by Eq. (10) reduces to:

$$\sigma_{SS} = \frac{1}{\alpha}\tilde{\tau}(\lambda_{SS})\tilde{G}(\lambda_{SS})\dot{\gamma}_{ss}. \tag{11}$$

On the other hand, the kinetic equation related to the TSR model given by Eq. (8) reduces to $\dot{\gamma}_{ss} = 1/\lambda_{SS}$, while that of given by Eq. (9) associated with VSR model leads to: $\dot{\gamma}_{Vss} = 1/\lambda_{SS}$. The subscript SS denotes the steady state. However, since in a limit of steady state, the elastic strain associated with the Maxwell model remains constant leading to $\dot{\gamma}_{Ess} = 0$, we have:

$$\dot{\gamma}_{ss} = \dot{\gamma}_{Vss} = 1/\lambda_{SS}. \tag{12}$$

Relationship between Eqs. (11) and (12) for a given value of $\lambda_{SS}$ results in the steady-state dependence of $\sigma_{SS}$ and $\dot{\gamma}_{ss}$, with a parameter $\mu$ and $\alpha$. In Fig 2, we plot $\sigma_{SS}$ versus $\dot{\gamma}_{ss}$ for three values of $\mu$ and $\alpha=1$. This figure is identical for both TSR and VSR models. For small values of $\mu$, the steady-state flow curve is monotonic in nature. However, beyond a specific value of $\mu$, which depends on the value of $g$ (in this work, we use $g=1$), the steady-state flow curve shows a non-monotonic behavior. The non-monotonic flow curve in a structural kinetic model suggests the presence of time-dependent yield stress [6, 7], and the system is expected to show shear banding for the low shear rates [29-33].

Fig. 3 shows a schematic of a flow field, which is applied to both the structural kinetic models. We assume that at $t=0$, the material is significantly aged such that the structure parameter is given by $\lambda = \lambda_i$. At $t=0$, the material in an undeformed state is subjected to a constant shear rate of $\dot{\gamma} = \dot{\gamma}_0$, while for $t \geq t_0$ the strain is kept constant ($\dot{\gamma} = 0$). In this work, we consider $\lambda_i = 100$ and $\dot{\gamma}_0 = 10$. The inset shows a schematic of the time (or $\lambda$) dependent Maxwell model. For a single-mode Maxwell model, the work done per unit volume on the same (in a dimensionless form) while deforming it from state 1 to 2 is given by:

$$E = E_v + E_e = \int_1^2 \sigma d\gamma_v + \int_1^2 \sigma d\gamma_e, \tag{13}$$

where $E_v$ is the energy dissipated in deforming the dashpot while $E_e$ is the energy required to deform the spring. In Eq. (13), the state 1 is typically that associated with $t=0$, as shown in Fig. 3, while state 2 can be any state at higher times. On the other hand, the dimensionless work per unit volume that can be recovered from the Maxwell model in a stretched state (state $1'$) to zero elastic strain state (state $2'$) having modulus $G(\lambda)$ at any value of $\lambda$ is:

$$W = \left| \int_{1'}^{2'} \sigma d\gamma_e \right| = \left| \int_{1'}^{2'} \tilde{G}(\lambda) \gamma_e d\gamma_e \right|. \tag{14}$$

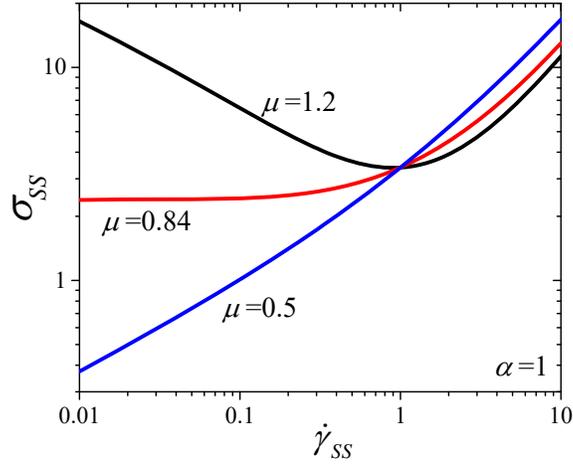

**Figure 2.** Steady-state flow curves for the structural kinetic model used in the present work for different values of $\mu$ and $\alpha = 1$. Change in $\alpha$ moves the flow curves vertically according to Eq. (11). This figure is identical for both TSR and VSR models.

We use Planck's statement of the second law [22] as discussed in the introduction to assess the validity of the same. Consequently, it is imperative that a thermodynamic cycle must get completed wherein the state of the Maxwell model after recovering the work is identical to that at time $t=0$ with respect to the initial value of elastic strain $\gamma_e$ and the initial value of structure parameter $\lambda_i$. In the present study, we start with a stress-free state, and hence at $t=0$, the elastic strain is $\gamma_e=0$. Since viscous strain induced in a system under constant $T$ and $P$ conditions does not change the free energy, attainment of the initial value, which is $\gamma_e=0$ and $\lambda = \lambda_i$, ensures completion of a thermodynamic cycle. Therefore, for those systems that allow a decrease in $\lambda$ under application of constant shear rate $\dot{\gamma} = \dot{\gamma}_0$, we apply constant strain for $t \geq t_0$ until the initial value of structure parameter $\lambda = \lambda_i$ is attained. Subsequently, we perform retraction of the spring associated with the Maxwell model in such a fashion that the structure parameter remains constant at $\lambda = \lambda_i$.

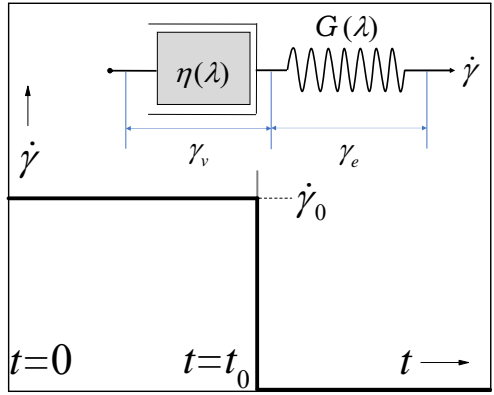

**Figure 3.** The schematic shows the flow field to which both the models are subjected to. At $t=0$, the shear rate of $\dot{\gamma}_0$ is applied to the time-dependent Maxwell model shown in the inset having an initial state ($\lambda = \lambda_i$). Subsequently at $t = t_0$, the strain is kept constant.

For the TSR model with $\lambda$ evolution given by Eq. (8), the process of contraction of the spring to recover the work can be performed by simply applying $\dot{\gamma} = 1/\beta\lambda_i$ (please note strain rate is applied in the opposite direction but since $\dot{\gamma}$ is the second invariant of the rate of strain tensor, it is always positive), which ensures no change in $\lambda$. On the other hand, for the VSR model with $\lambda$ evolution given by Eq. (9), $\lambda$ remains constant if an instantaneous retraction of spring is carried out. This is because during an instantaneous retraction process, the viscous flow does not get induced $(\dot{\gamma}_v = 0)$. Consequently, in both the cases $\tilde{G}(\lambda)$ remains constant, and the work recovered is given by:

$$W = \frac{1}{2}\tilde{G}(\lambda)\gamma_e^2. \tag{15}$$

Interestingly for the VSR model, wherein $\lambda$ evolution given by Eq. (9), Eq. (15) represents the maximum work that can be recovered for a given value of $\lambda$ and $\gamma_e$. This is because any process of strain reversal of spring slower than instantaneous retraction will induce nonzero $\dot{\gamma}_v$ that will cause a decrease in $\lambda$ according to Eq. (9). Although one can age the system under quiescent conditions after complete retraction to obtain $\lambda = \lambda_i$, the work recovered will always be less than that given by Eq. (15) as a decrease in $\lambda$ will cause a decrease in $\tilde{G}(\lambda)$. For the TSR model for which $\lambda$ evolution given by Eq. (8), Eq. (15) does not lead to the maximum work, however that does not affect the conclusion as long as the thermodynamic cycle is complete as discussed below.

The rest of this section is arranged as follows. In the following subsection, we carry out energy analysis of the proposed structural kinetic models with both the types of structure factor evolution expressions Eqs. (8) and (9) for which we take time $t_0$ to be that time at which $\lambda=1$ unless mentioned otherwise. In the subsequent subsection, we analyze non-monotonic stress relaxation associated with the proposed VSR model (with change in expression for relaxation time dependence on $\lambda$), wherein we consider time $t_0$ to be that associated with just acquiring the steady state.

### A.  Energy analysis

To begin with, we consider the TSR model with the structure factor evolution equation given by Eq. (8), wherein the total strain rate ($\dot{\gamma}$) is responsible for causing the rejuvenation. We first consider the case of $\alpha=0$, which represents the dashpot of the Maxwell model having infinite viscosity. Consequently, this limit ($\alpha=0$) represents thixoelastic material. For this special case, Eqs. (8) and (10) are solved along with the dimensionless modulus given by: $\tilde{G}(\lambda) = \left[1 + g\ln\left(1 + \lambda^\mu\right)\right]$. In Fig. 4(a), we plot the evolution of $\lambda$, $\sigma$ and $\gamma_e$ under application $\dot{\gamma}_0 = 10$ as a function of time, wherein the vertical gray line represents $t = t_0$. It can be seen that for $t < t_0$, $\lambda$ decreases with time and the corresponding stress first increases, shows a maximum, and then decreases. Since, for this specific case, stress is a product of modulus and elastic strain, initially an increase in strain enhances the stress, but eventually decrease in $\lambda$ shows effect and stress displays an overshoot. Since in the studied case $\alpha=0$, the total strain is equal to the elastic strain that, expectedly, shows a linear increase with time for $t < t_0$. For $t \geq t_0$, strain is kept constant ($\dot{\gamma} = 0$), $\lambda$ increases linearly with time, causing modulus and hence stress to increase with time.

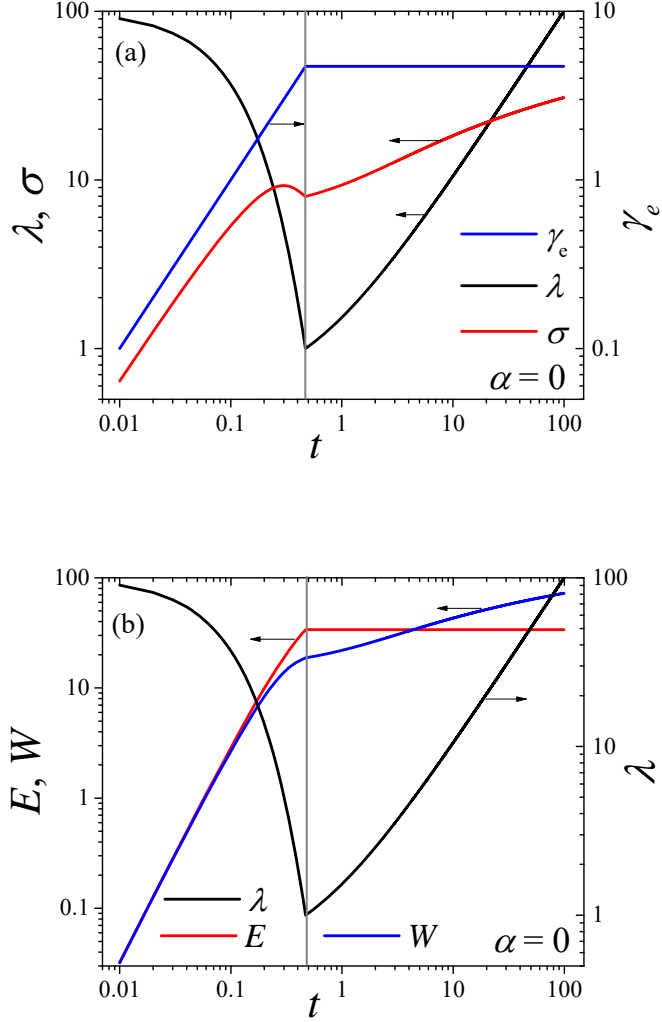

**Figure 4.** Results of the TSR model for which $\lambda$ evolution expression given by Eq. (8), where rejuvenation is induced by the total shear rate for a flow field shown in Fig 3 with $\alpha = 0$ (only elastic spring is present in the Maxwell model in a limit $\eta \to \infty$). Evolution of (a) $\lambda$, $\sigma$, $\gamma_e$ and (b) $\lambda$, $E$, $W$ are plotted as a function of time. The vertical grey line represents $t = t_0$.

In Fig. 4(b), we plot total energy ($E$) provided to the model to undergo mentioned deformation as well as instantaneous recoverable work ($W$) as a function of time. It can be seen that for $t < t_0$, both $E$ and $W$ increase with time owing to an increase in strain with time. However, since $\lambda$ decreases with time in this region (causing a decrease in $\tilde{G}(\lambda)$ with time), the difference between $E$ and $W$ goes on increasing, with $E$ always greater than $W$. For $t \geq t_0$, strain remains constant, and surrounding does not perform any more work on the system leading to constant $E$. However since $\dot{\gamma} = 0$, according to Eq. (8), $\lambda$ increases linearly with time, which causes an enhancement in $\tilde{G}(\lambda)$. Consequently, the work that can be recovered from the system ($W$) increases rapidly according to Eq. (15) and as shown in Fig. 4(b). Eventually $W$ crosses $E$ and continues to increase further. At a certain time, $\lambda$ reaches the initial value. If at this state, work is recovered from the spring as represented by Eq. (15), the system will reach its initial state in terms of strain as well as $\lambda$. However, at this stage, since recovered work is greater than the work done on the system during the strain cycle, wherein the system attains the initial state, the second law of thermodynamics gets violated. Such possibility of violation of the second law was predicted by Larson [10]. Accordingly, he rejected the existence of the ideal thixo-elastic material that follows the evolution equation given by Eq. (8).

It is important to note that, as discussed before, in a rejuvenated (structureless) state of a material, $\lambda$ undergoes spontaneous increase as a function of time under quiescent conditions. Consequently, free energy is a monotonically decreasing function of $\lambda$. Therefore, a mere higher value of extracted work than work done on the system will not necessarily violate the second law. As clearly mentioned in Planck's statement, the cycle must get completed, wherein the initial state of the free energy (initial value of $\lambda$) must be achieved, and the corresponding extracted work needs to be greater than the work done on the system for the second law to get violated. As mentioned before, work recovered represented by Eq. (15), is not the maximum work when the $\lambda$ evolution equation is given by Eq. (8), as strain reversal can be carried out more slowly such that $\lambda$ (and hence $\tilde{G}(\lambda)$) increases. The high $\lambda$ state of such a system can be rejuvenated to the initial value $\lambda_i$ by subjecting it to oscillatory shear with large frequency but small strain amplitude that costs only minimal work. However, since even with Eq. (15) the second law gets violated, the inference of the analysis does not change.

In Fig. 4, we consider a particular case of thixo-elastic material in a framework of the TSR model, wherein the viscosity of dashpot shown in the inset of Fig. 3 is infinite ($\alpha = 0$). Next, we study a case of the TSR model, wherein viscosity or the relaxation time of the Maxwell model is still very high (but not infinity) such that $\alpha = 0.1$. In Fig. 5(a), we plot the evolution of $\lambda$, $\sigma$, $\gamma_e$ and $\gamma_v$ under application $\dot{\gamma}_0 = 10$ as a function of time. As shown in the figure, for $t < t_0$, $\lambda$ decreases with time while the corresponding stress shows an overshoot. Moreover, with an increase in time, both $\gamma_e$ and $\gamma_v$ increase; but, owing to the high viscosity of dashpot, increase in $\gamma_e$ is far more stronger than that of $\gamma_v$. For $t \geq t_0$, the cumulative strain in spring as well as dashpot is kept constant ($\dot{\gamma} = 0$), and hence $\lambda$ increases linearly with time leading to enhancement in modulus as per Eq. (7). Furthermore, for $t \geq t_0$, as expected $\gamma_e$ decreases with time at the cost of an increase in $\gamma_v$, so that their sum remains constant as shown in Fig 5(a). There are two factors contributing to the stress, increase in the modulus and decrease in $\gamma_e$. As a result of the high viscosity of dashpot, an increase in the modulus dominates, causing stress to increase with time.

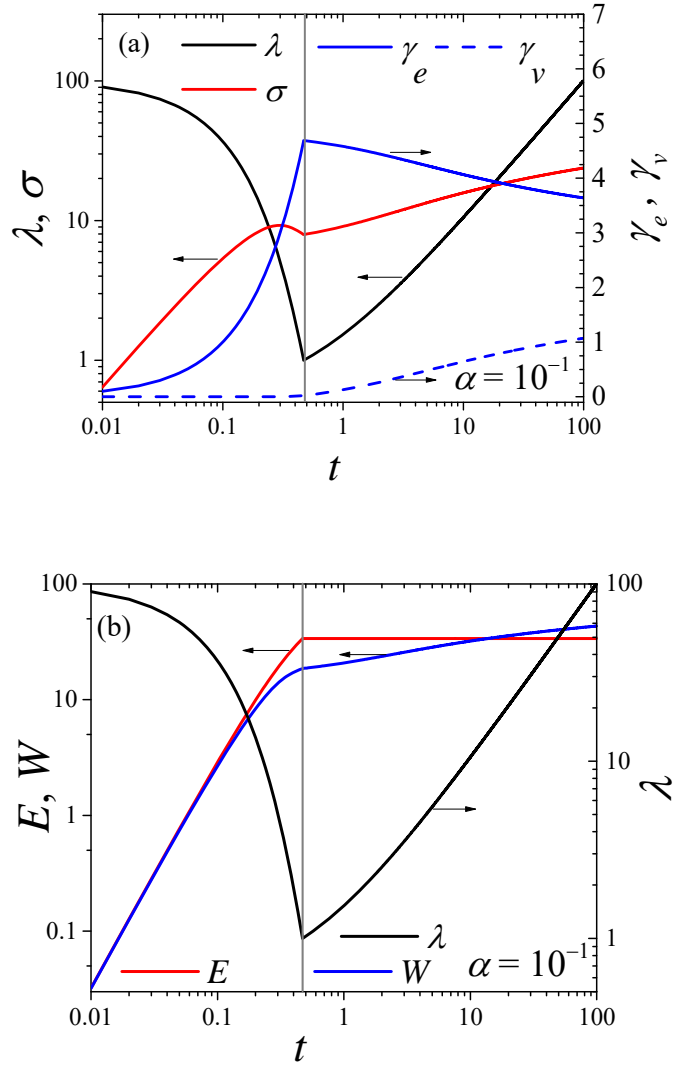

**Figure 5.** Results of the TSR model for which $\lambda$ evolution expression given by Eq. (8), where rejuvenation is induced by the total shear rate for a flow field shown in Fig 3 with $\alpha = 0.1$. Evolution of (a) $\lambda$, $\sigma$, $\gamma_e$, $\gamma_v$ and (b) $\lambda$, $E$, $W$ are plotted as a function of time. The vertical grey line represents $t = t_0$.

In Fig. 5(b) we plot $E$, $W$ and $\lambda$ as a function of time. It can be seen that $E$ (work done by the surrounding on the system) increases for $t < t_0$ and remains constant for higher times. For $t \geq t_0$, since the total strain remains constant, $E$ also remains constant. However, in this regime, $\lambda$ as well as modulus increase with time, causing an increase in $W$ despite decrease in $\gamma_e$. Eventually, at a certain time, $W$ increases beyond $E$. Ultimately, $\lambda$ reaches the initial value, and at that state, $W$ is greater than $E$, suggesting violation of the second law. It should be noted that in this cycle, the $\lambda$ as well as $\gamma_e$ reach the initial state (initial and final value of $\gamma_e = 0$, as work is extracted by reversing the deformation completely while keeping $\lambda$ constant). However, the final value of the viscous strain may not be the same as the initial value of the viscous strain. Nevertheless, as discussed before, this does not prohibit consideration of completion of the cycle as under constant $T$ and $P$ conditions, the viscous strain does not contribute to the free energy. Example of Fig. 5 for $\alpha = 0.1$ suggests that thixo-viscoelastic material also may show a violation of the second law if the evolution of $\lambda$ considers rejuvenation to occur by the total shear rate $(\dot{\gamma})$ as considered in TSR model represented by Eq. (8).

We next consider the case of TSR model for $\alpha = 1$ that represents further lower viscosity of the dashpot (or lower relaxation time of the system) compared to what is considered in Figs. 4 and 5. Under application of $\dot{\gamma}_0 = 10$, the time evolution of $\lambda$, $\sigma$, $\gamma_e$ and $\gamma_v$ is plotted in Fig. 6(a). It can be seen that, $\lambda$ decreases with time while $\sigma$ shows an overshoot for $t < t_0$. Additionally, both $\gamma_e$ and $\gamma_v$ increase with time; however, the increase in $\gamma_e$ is much stronger than that of $\gamma_v$. This is because, although lower than that considered for the case of $\alpha = 0.1$ shown in Fig 5, the viscosity associated with $\alpha = 1$ is still high enough to cause significant deformation of spring than dashpot. For $t \geq t_0$, wherein the total strain is kept constant ($\dot{\gamma} = 0$), and $\lambda$ increases linearly with time, causing an increase in modulus as per Eq. (7). In this regime, $\gamma_e$ decreases while $\gamma_v$ increases keeping their sum constant. However, compared to Fig 5(a), the change in both strains is very significant. This leads to continuous relaxation of stress despite an increase in modulus (due to an increase in $\lambda$) as shown in Fig 6(a).

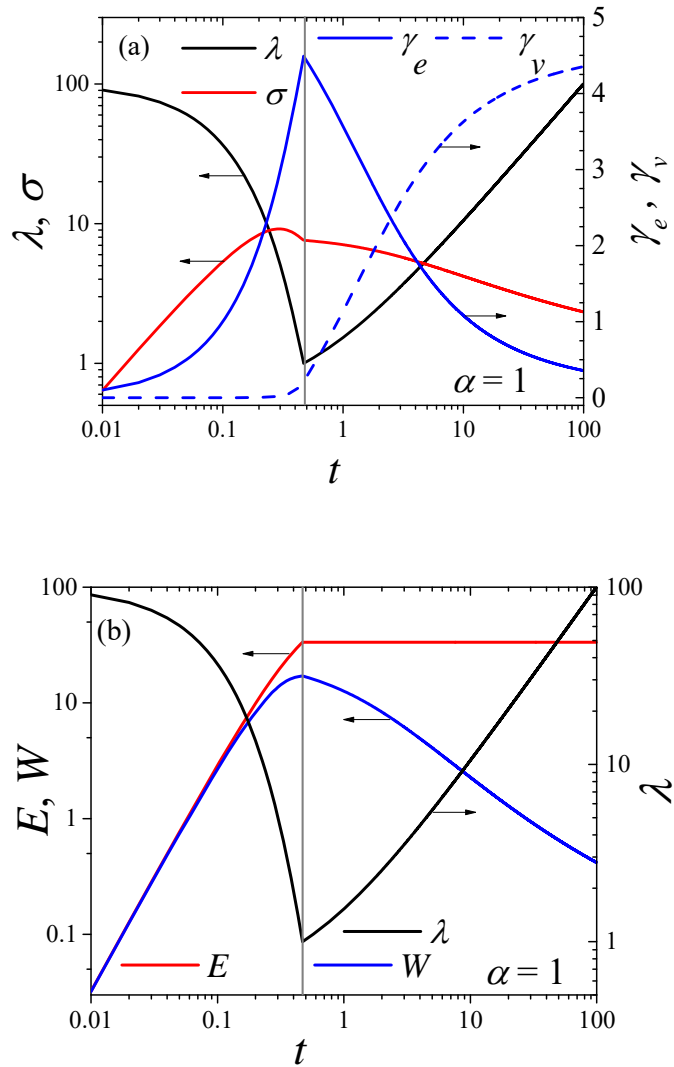

**Figure 6.** Results of the TSR model for which $\lambda$ evolution expression given by Eq. (8), where rejuvenation is induced by the total shear rate for a flow field shown in Fig 3 with $\alpha = 1$. Evolution of (a) $\lambda$, $\sigma$, $\gamma_e$, $\gamma_v$ and (b) $\lambda$, $E$, $W$ are plotted as a function of time. The vertical grey line represents $t = t_0$.

In Fig. 6(b), we plot temporal evolution of $E$, $W$ and $\lambda$ for the case of $\alpha = 1$. It can be seen that for $t < t_0$, $E$ increase, while for $t \geq t_0$, $E$ remains constant as the system as a whole does not undergo any deformation. As expected, for $t < t_0$, the recoverable work $W$, given by Eq. (15), is always lesser than $E$ throughout the time domain. However for $t \geq t_0$, although $\lambda$ as well as modulus increase with time, a significant decrease in $\gamma_e$, owing to lower viscosity of dashpot causes a continuous decrease in $W$. Eventually, $\lambda$ reaches the initial value but $W$ is significantly smaller than $E$. Therefore, no violation of the second law occurs for the studied case of low viscosity of the dashpot.

In order to understand the effect of $\alpha$ and $\mu$ on violation of the second law for the TSR model for the studied flow field and the expression of $W$ given by Eq. (15), we plot $W/E$ at the point of completion of the thermodynamic cycle (when the initial value of $\lambda$ is achieved during constant strain step) as a function of $\mu$ for different values of $\alpha$ in Fig 7. Since $W/E > 1$ represents a recovery of more work that whet is done on the system at the end of the cycle, it suggests a violation of the second law. In general, It can be seen that, $W/E$ increases with for increase in $\mu$ and decrease in $\alpha$. For $\alpha = 0$ and for the non-zero values of $\mu$ ($\mu = 0$ represents non-aging equilibrium material), we observe $W/E > 1$ suggesting clear violation of the second law, emphasizing ideal thixoelastic material is indeed unrealistic. For higher values of $\alpha$, violation of the second law depends on the value of $\mu$. Overall the TSR formulation represnted by $\lambda$ - evolution given by Eq. (8) in a viscoelastic material has an inherent weakness of possible violation of the second law.

The analysis of the TSR model reported in Figs 4 to 6 consider rejuvenation in the structural kinetic model is caused by the total shear rate as per Eq. (8). We now study the results of the otherwise identical structural kinetic model except that the evolution expression for $\lambda$ is given by Eq. (9), wherein rejuvenation is produced by the viscous component of the shear rate ($\dot{\gamma}_v$). This model has been represented as the VSR model. We first analyze a system with $\alpha = 0$. This case indicates the viscosity of the dashpot to be infinity, and the Maxwell model comprises only spring with time-dependent modulus, and hence, $\dot{\gamma}_v = 0$. Consequently, $\lambda$ increases linearly as a function of time irrespective of the value of $\dot{\gamma}$. Therefore, for this specific case, we take $t_0 = 10$, such that for $t < t_0$ the model is subjected to $\dot{\gamma}_0 = 10$ while for $t \geq t_0$ the strain is kept constant ($\dot{\gamma} = 0$). In Fig. 8(a), we plot the evolution of $\lambda$, $\sigma$ and $\gamma_e$ as a function of time. It can be seen that for $t < t_0$, both $\lambda$ and $\sigma$ increase with time.

Furthermore, as expected, the elastic strain, which is equal to the total strain, shows a linear increase with time for $t < t_0$. For $t \geq t_0$, the strain is kept constant ($\dot{\gamma} = 0$) and because there is no dashpot, $\lambda$ continues to increase linearly with time, causing modulus to increase. However, although strain is constant, since the stress is a product of modulus and elastic strain, it can be seen to be increasing with time.

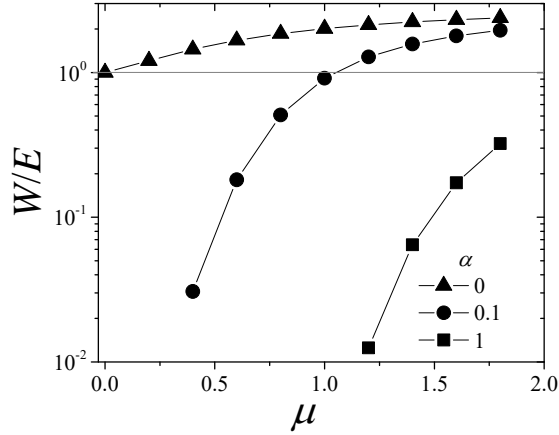

**Figure 7.** The ratio of the work recovered from the to the system to the energy input $\left(W/E\right)$ at the point of completion of cycle (when the original value of $\lambda$ is achieved as per Figure 3) is plotted as a function of $\mu$ for different values of $\alpha$ for the TSR model for which $\lambda$ evolution expression given by Eq. (8). As per Planck's statement, the second law gets violated for $W/E > 1$. The lines are a guide to the eye.

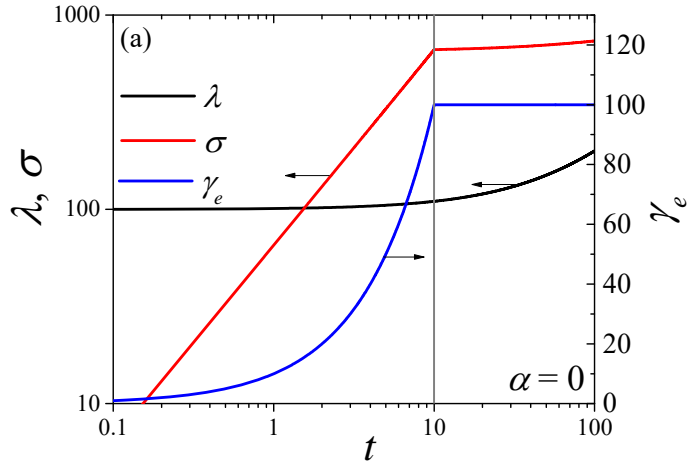

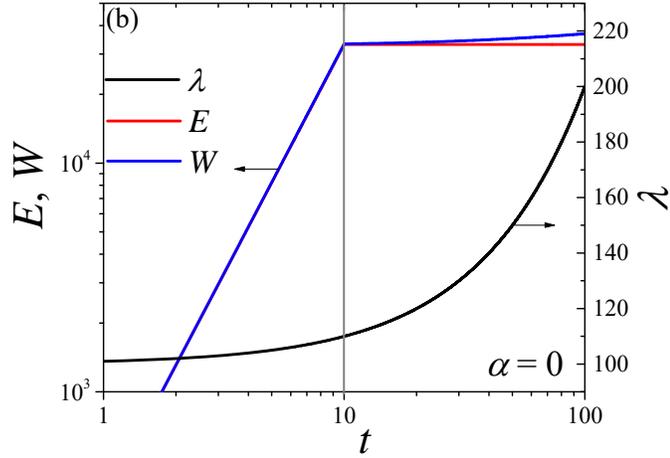

**Figure 8.** Results of the VSR model for which $\lambda$ evolution expression given by Eq. (9) , where rejuvenation is induced by the shear rate associated with dashpot ($\dot{\gamma}_v$) for a flow field shown in Fig 3 with $\alpha = 0$. Evolution of (a) $\lambda$, $\sigma$, $\gamma_e$ and (b) $\lambda$, $E$, $W$ are plotted as a function of time. The vertical grey line represents $t = t_0$.

In Fig. 8(b), we plot $E$, $W$, and $\lambda$ as a function of time. As expected, $E$ and $W$ both increase with time. However, since modulus continuously goes on increasing, $W$ is always larger than $E$. For $t \geq t_0$, $E$ remains constant, but again since $\lambda$ continuously goes on increasing with time causing an increase in the modulus, $W$ increases with time. Therefore, any time during this process, the work extracted will always be greater than the work performed on the system. However, since material continuously undergoes structural buildup, as described by an increase in $\lambda$, which is equivalent to a decrease in free energy, the additional work obtained from the system is due to the lowering of the free energy. Since the system does not perform a cycle during this process, as necessitated by Planck's statement, the question of violation of the second law does not arise. It should be noted that this case is primarily academic as rarely any soft material that ages acquires only the elastic strain under application of prolonged finite shear rate. Nonetheless, it is a limiting case with $\alpha = 0$ and hence deserves discussion. We return to this case in the next section. It is tempting to call the structural kinetic model with evolution expression for $\lambda$ given by Eq. (9) for $\alpha = 0$ an ideal thixo-elastic model. However, in this case while modulus increases with time, since there is no provision for it to decrease, it cannot be called as a thixo-elastic model.

Therefore, Larson's [10] proposal that ideal thixo-elastic materials do not exist remains valid.

In Fig. 9, we plot the results associated with the VSR model, wherein the evolution of structure parameter considers rejuvenation due to viscous shear rate given by Eq. (9) for $\alpha = 0.01$. In Fig. 9(a), we plot the evolution of $\lambda$, $\sigma$, $\gamma_e$ and $\gamma_v$ as a function of time under application of $\dot{\gamma}_0 = 10$. At lower times $\gamma_e$ shows a significant increase compared to $\gamma_v$, therefore $\dot{\gamma}_v << \dot{\gamma}$ leading to $\lambda \dot{\gamma}_v < 1$ ($\lambda_i = 100$). Such behavior, according to Eq. (9) results in an increase in $\lambda$ as a function of time. However, as time progresses, the rate of viscous strain increases with time so that $\lambda \dot{\gamma}_v$ increases above unity, and eventually $\lambda$ decreases and reaches a value of unity. Subsequently, the strain is kept constant ($t \geq t_0$). Although the total strain rate is zero, the viscous strain rate remains high. Therefore, for a very short period of time beyond $t = t_0$, $\lambda$ continues to decrease, reaches a value close to 0.1, and then starts increasing again as $\lambda \dot{\gamma}_v$ becomes less than unity. The stress for $t < t_0$ goes on increasing, and at the point of decrease in $\lambda$, undergoes a sharp decrease, thereby showing an overshoot. For $t \geq t_0$, $\gamma_e$ decreases at the cost of $\gamma_v$ keeping the total strain constant. Interestingly, $\lambda$ increases in this region, causing an increase in modulus while $\gamma_e$ decreases. Consequently, the stress continues to decrease for a certain period, which is dominated by a decrease in $\gamma_e$. However, beyond a particular point, a decrease in $\gamma_e$ (as well as increase in $\gamma_v$) slows down. This enhances the rate of increase in $\lambda$ and hence the modulus. As a result, the enhancement in modulus dominates, and the stress starts increasing with time, thereby showing a non-monotonic relaxation of stress under the application of constant strain.

In Fig. 9(b), $E$, $W$, and $\lambda$ are plotted as a function of time. As expected, $E$ and $W$ increase with time. However, as $\lambda$ decreases, $W$ starts decreasing as a function of time owing to a decrease in modulus. For $t \geq t_0$, $E$ remains constant as total strain is constant but owing to relaxation of spring $E_e$ is expected to decrease. The recoverable work $W$ is proportional to modulus and $\gamma_e^2$. For $t \geq t_0$, $\gamma_e$ decreases with time while $\lambda$ goes on increasing with time causing an increase in modulus. So, during the initial part of the relaxation, the same as what is observed for the stress, $W$ decreases with time. However, as a decrease in $\gamma_e$ retards, and $\lambda$ (and hence modulus) shows sharp increase, $W$ increases, thereby showing a minimum. Nevertheless, $W$ always remains smaller than $E$, including the point where $\lambda$ attains the initial value. Consequently, the second law of thermodynamics is not violated.

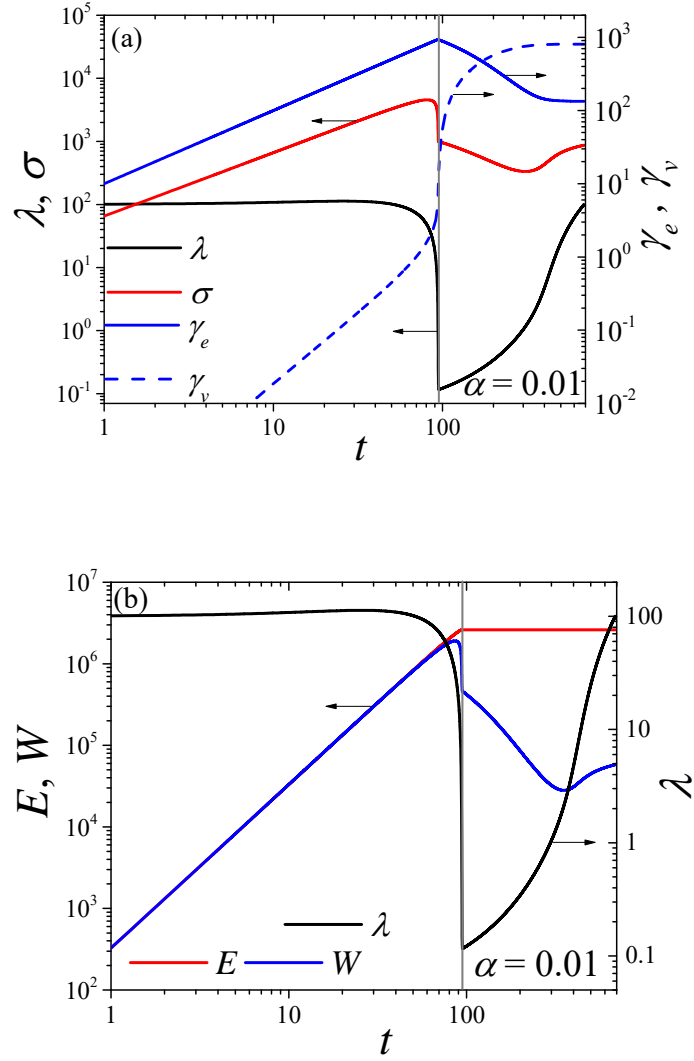

**Figure 9**. Results of the VSR model for which $\lambda$ evolution expression given by Eq. (9), where rejuvenation is induced by the shear rate associated with dashpot ($\dot{\gamma}_v$) for a flow field shown in Fig 3 with $\alpha = 0.01$. Evolution of (a) $\lambda$, $\sigma$, $\gamma_e$, $\gamma_v$ and (b) $\lambda$, $E$, $W$ are plotted as a function of time. The vertical grey line represents $t = t_0$.

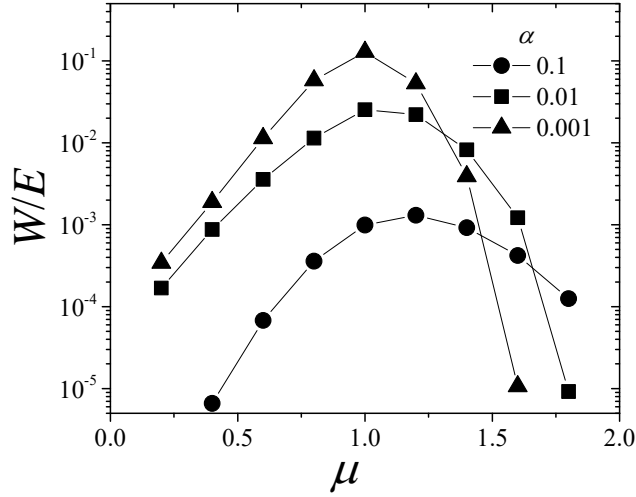

**Figure 10.** The ratio of the work recovered from the to the system to the energy input $\left(W/E\right)$ at the point of completion of the cycle (when the original value of $\lambda$ is achieved as per Figure 3) is plotted as a function of $\mu$ for different values of $\alpha$ for the VSR model for which $\lambda$ evolution expression given by Eq. (9). As per Planck's statement, the second law gets violated for $W/E > 1$. The lines are guide to the eye.

We also investigate an effect of $\alpha$ and $\mu$ on $W/E$ at the point of completion of thermodynamic cycle (when the initial value of $\lambda$ is achieved during constant strain step) for 1-510, we plot $W/E$ at the point of completion of the thermodynamic cycle as a function of $\mu$ for different values of $\alpha$. Contrary to what was observed for the TSR model, $W/E$ shows a non-monotonic dependence on $\mu$ for the VSR model. Furthermore, for lower values of $\mu$, a decrease in $\alpha$ causes increase in $W/E$. However, after the maximum, a decrease in $W/E$ becomes sharper with a decrease in $\alpha$. Such behavior primarily occurs because of the following dependencies of $E$ and $W$ on $\alpha$ and $\mu$. Firstly $E$ increases with increase in $\mu$ such that enhancement becomes stronger at higher values of $\mu$ and lower values of $\alpha$. On the other hand, an increase in $W$ gets weaker with an increase in $\mu$ leading to the observed behavior. To further understand behavior shown in Figs. 9 and 10, let us consider a case wherein the viscosity of the dashpot is suitably large ($\alpha$ is sufficiently small) such that $\dot{\gamma}_v$ for most of the time domain $t < t_0$ will be so small that $\lambda$ is expected to be higher than $\lambda_i$ before suddenly decreasing to any pre-assigned value. Consequently, the work done on the system would be large. For $t \geq t_0$, on the other hand, owing to the high viscosity of the dashpot, $\lambda$ is expected to increase at a faster rate to reach $\lambda_i$, and hence work recovered will never be greater than that of work done. For higher values of $\alpha$, dissipation associated with the dashpot will not allow the recovered work to be more than work done on the system, as shown in Fig 10. Consequently, with $\lambda$ evolution given by Eq. (9), the second law will never get violated irrespective of the value of $\alpha$. It is also interesting to note that mere increase in $\mu$, which causes greater enhancement in modulus as a function of time, does not cause violation of the second law. As discussed above, this is simply because an increase in $\mu$ at higher values causes $E$ to increase at much greater strength than that of $W$.

It should be mentioned that, although we use evolution equation proposed by Coussot and co-workers [6, 26] as represented by Eq. (3), the model proposed by them is for an inelastic fluid. Consequently, since there is no consideration of modulus as well as relaxation time, Coussot and co-workers' model [6, 26] must show instantaneous stress relaxation. In any inelastic constitutive equation, wherein the total shear rate is also viscous shear rate, the second law of thermodynamics must always get validated. On a broader note, the analysis presented in this work is not limited to any particular model but should be applicable for any structural kinetic model, wherein rejuvenation is due to the strain rate associated with the dissipative processes.

### B. Non-monotonic stress relaxation

For a time-dependent Maxwell model, the stress is a product of modulus and strain in the spring. During stress relaxation, the total strain associated with the Maxwell model is held constant. Consequently, strain in the spring decreases at the cost of strain in the dashpot with an increase in time. However, under application of constant strain, given that viscosity of dashpot is not too low so that $\lambda \dot{\gamma}_v < 1$, $\lambda$ (and hence the modulus) increases with time. This behavior eventually produces an increase in stress leading to non-monotonic stress relaxation, as shown in Fig. 8(a). As discussed in the introduction section, Hendricks and coworkers [12] reported non-monotonic stress relaxation in three experimental systems. Their data associated with one of the systems is shown in Fig. 1. The major difference between the trend of the non-monotonic stress relaxation data of Hendricks and coworkers [12] and that of shown in Fig. 8(a) is complete relaxation of the experimental systems in the former. This is because the system described by the structural kinetic model (represented by Eqs. (6), (7), (9) and (10)) does not attain equilibrium. To address this issue, we make the following amends to the model so that it describes a system undergoing complete relaxation.

In the structural kinetic model proposed in section II, we consider $\lambda$ to increase indefinitely and the corresponding relaxation time and modulus are given by Eqs. (6) and (7), which also show an indefinite increase with time as observed for many systems that do not attain equilibrium and continue to undergo physical aging over practically explorable timescales [13, 15]. The systems studied by Hendricks and coworkers [12], on the other hand, show structure formation with time after cessation of shear flow. However, these systems eventually attain equilibrium as suggested by the complete relaxation of the stress. To model this behavior, we consider the $\lambda$-evolution expression given by Eq. (9) but propose relations for relaxation time and modulus given by:

$$\tau = \tau_0 \tilde{\tau}(\lambda) = \tau_0 \left[1 + \left[\lambda_t \left(1 - \exp(\lambda/\lambda_t)\right)\right]^\mu \right] \text{ and} \qquad (16)$$

$$G = G_0 \tilde{G}(\lambda) = G_0 \left[1 + g \ln(\tilde{\tau})\right] = G_0 \left[1 + g \ln\left[1 + \left[\lambda_t \left(1 - \exp(\lambda/\lambda_t)\right)\right]^\mu\right]\right], \qquad (17)$$

where $\lambda_t$ is the threshold value of structure parameter such that for $\lambda/\lambda_t \gg 1$, relaxation time as well modulus become constant. Eqs. (9), (16) and (17), therefore, suggest that for $\lambda \gg \lambda_t$ the structural buildup occurring in a material is negligible, and for all practical purposes, equilibrium can be assumed in such a limit. In the inset of Fig. 11(a), we plot $\tilde{\tau}$ and $\tilde{G}$ as a function of $\lambda$ for $\lambda_t = 100$. It can be seen that for $\lambda \gg \lambda_t$, both $\tilde{\tau}$ and $\tilde{G}$ indeed reach a constant value.

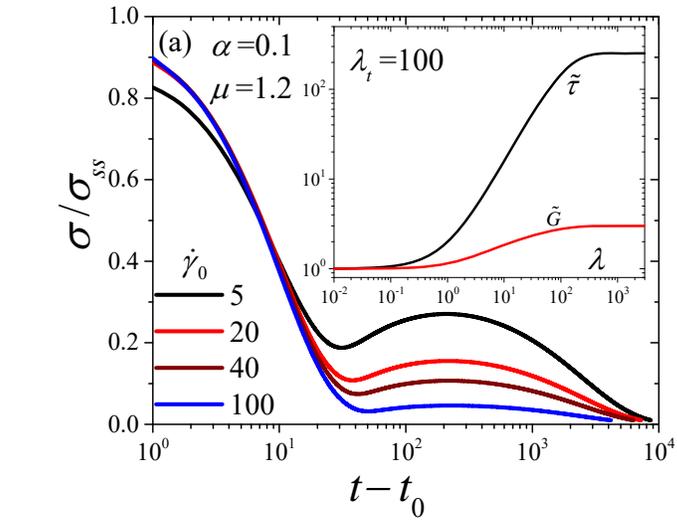

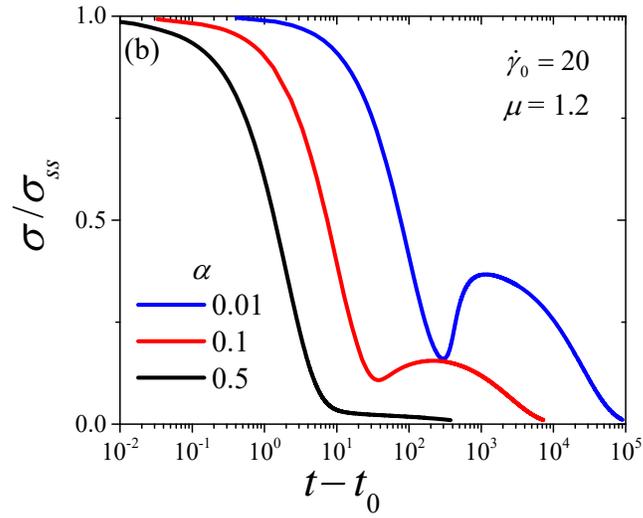

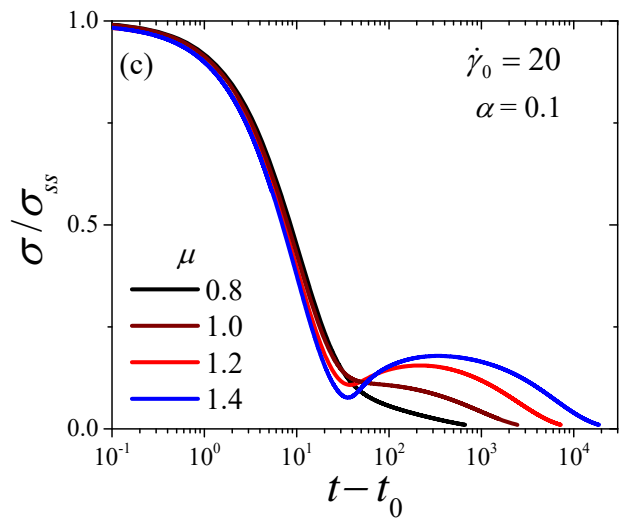

**Figure 11.** Stress relaxation after stoppage of steady-state ($\dot{\gamma}_0$) shear under constant strain conditions is plotted as a function of time for the amended VSR model discussed in section III.B. Stress is normalized by shear stress at the steady-state ($\sigma_{ss}$). (a) The stress relaxation is plotted at different steady state shear rates as mentioned in the legend for $\alpha = 0.1$ and $\mu = 1.2$. The inset shows the evolution of $\tilde{\tau}(\lambda)$ and $\tilde{G}(\lambda)$ as described by Eqs. (16) and (17). In (b), the stress relaxation is plotted for different values of $\alpha$ at $\dot{\gamma}_0 = 20$ and $\mu = 1.2$ ($\alpha$ decreases from the left to right), while in (c) the stress relaxation is plotted for different values of $\mu$ at $\dot{\gamma}_0 = 20$ and $\alpha = 0.1$ (on the right hand side, $\alpha$ decreases from top to bottom).

We solve the amended VSR model given by Eqs. (9), (10), (16) and (17) with $\alpha = 0.1$ for a flow field given by Fig. 3 wherein we subject the model to a constant shear rate of $\dot{\gamma}_0$ for a time $t_0$, which is the time associated with the attainment of steady-state. Subsequently, we put $\dot{\gamma} = 0$ and obtain the evolution of $\sigma$ as a function of time. In Fig. 11(a), we plot stress normalized by steady stress ($\sigma/\sigma_{ss}$) as a function of $t - t_0$ for different values of $\dot{\gamma}_0$. It can be seen that there is an astonishing similarity between the experimental data shown in Fig. 1 and the results of the amended VSR model shown in Fig 11, wherein stress first relaxes followed by a noticeable increase and subsequent decrease to attain the equilibrium state. With an increase in $\dot{\gamma}_0$ the normalized stress can be seen to be decreasing to lower values, as also observed experimentally. In Fig. 11(b) and (c) we respectively assess effect of $\alpha$ and $\mu$ on the non-monotonic stress relaxation behavior. It can be seen that decrease in $\alpha$ (increase in viscosity of dashpot) and an increase in $\mu$ lead to more pronounced non-monotonicity. We also confirmed that anytime during relaxation of stress, including that part where stress increases with time, the recoverable work is always less than the work done on the system. This behavior is equivalent to that shown in Fig. 9(b). This suggests that the non-monotonic stress relaxation may not violate the second law of thermodynamics.

## IV. Analysis of the results

The primary result of the present work is how the soft glassy (thixotropic) materials get rejuvenated. We show that the second law of thermodynamics is not violated when a structural kinetic model considers only the viscous shear rate to induce rejuvenation. Interestingly, various theories on the rheology of soft glassy materials indeed propose rejuvenation to get augmented with an increase in the stored elastic energy. In the SGR formulation [14], the depth of energy well associated with an arrested element gets reduced by the elastic energy acquired by the same. An arrested element must overcome this barrier in order to diffuse out of the energy well (a local rejuvenation event). Therefore, the more is the elastic energy, the more probable is a local rejuvenation event. In the present formulation, given by the VSR model, while it appears that the rejuvenation is solely due to the viscous rate of strain, the elastic strain associated with the spring does play an important role. In a Maxwell model, the more the strain induced in the spring, the more is the stress, and consequently, more is the viscous rate of strain. Therefore, the VSR model does implicitly capture the underlying physics of elastic stored energy, augmenting the process of rejuvenation.

We next discuss physical aging under strained and unstrained states by a schematic shown in Fig. 12. Consider a structurally arrested particle associated with a soft glassy material. The structural arrest is possible either through the physical bond formation between the particles (gel structure) or through the crowding of particles (glass structure). We show a schematic of a section of a gel structure (Fig. 12 top left), wherein the particle under consideration is arrested due to attractive physical interactions (bonds) it shares with the neighboring particles. Irrespective of the nature of microstructure, the structural arrest can be represented by the particle residing in an energy well (Fig. 12 middle left). Under the quiescent condition, the particle undergoes microstructural rearrangement and chooses those states that take it to progressively lower well depths (Fig. 12 bottom left). This enhanced bond energy (or well depth $U$) not just increases the relaxation time (as per $\tau = \tau' \exp(U/kT)$) but also the modulus $\left(G \approx U/b^3\right)$ [28]. Application of strain field (Fig 12, top right) at any point in time enhances the elastic energy of the system (Fig 12, middle right). If material is held under the application of constant strain (Fig 12, bottom right), by virtue of its thermal energy, the particle may still undergo microscopic rearrangements by which it can lower its energy as a function of time. In other words, the strengthening of the interparticle bond may occur even under the strained state. Strengthening of a bond that is under tension (owing to finite strain) may cause an increase in the well depth and hence enhancement in the modulus. However, since the strain is constant, it can lead to an increase in stress. Such enhancement of modulus (and therefore stress) is solely at the cost of free energy of the system, and hence it does not violate the second law. As discussed in the introduction, under constant $T$ and $P$ conditions, in the process of physical aging, the Gibbs free energy reduces as a function of time. Under quiescent (stress- free) conditions with constant $T$ and $P$, thermodynamically, the lowering of free energy (specific Gibbs free energy $f$) with time can simply be represented as:

$$\frac{df}{dt} = -\dot{q}_{out}, \tag{18}$$

where $\dot{q}_{out}$ is the rate of heat transfer from the system to the surrounding. We are also assuming that the volume of the system changes negligibly, so a change in Gibbs free energy is also equal to a change in Helmholtz free energy. To have validation of the second law, $\dot{q}_{out}$ must always be positive [34].

Let us now discuss a limiting case of a material wherein viscosity is so high that it only undergoes elastic deformation. We discussed the behavior of such a system in Fig. 7 and deduced that although the recovered work is greater than the applied work, the second law does not get violated as a thermodynamic cycle does not get completed as required by Planck's statement of the second law. Let us consider a material in a semi-solid state that is undergoing physical aging. At a particular time, let the specific Gibbs free energy be $f_i$. At that time, we subject it to step strain, so that $w_i$ is the work done on it. Let us assume that this work $w_i$ is entirely elastic as the material does not get rejuvenated at all (no viscous flow). After a certain time $\Delta t$, let the Gibbs free energy be $f_f$, such that $f_f < f_i$. Owing to physical aging, since modulus has increased, the work recovered after time $\Delta t$ be $w_f$ (such that $w_f > w_i$, and hence net work output is $w_{net} = w_f - w_i$). We can therefore write the energy balance as:

$$f_f - f_i = -w_{net} - q'_{out}, \tag{19}$$

where $q'_{out}$ is the heat transferred from the system to the surrounding and is always positive. If the exact extent of physical aging (decrease in Gibbs free energy from $f_i$ to $f_f$) is to be carried out under the quiescent conditions such that $q_{out}$ heat transfers from the system to the surrounding, we have:

$$f_f - f_i = -q_{out} \qquad (20)$$

Comparing Eq. (19) and (20) leads to:

$$w_{net} = q_{out} - q'_{out}. \qquad (21)$$

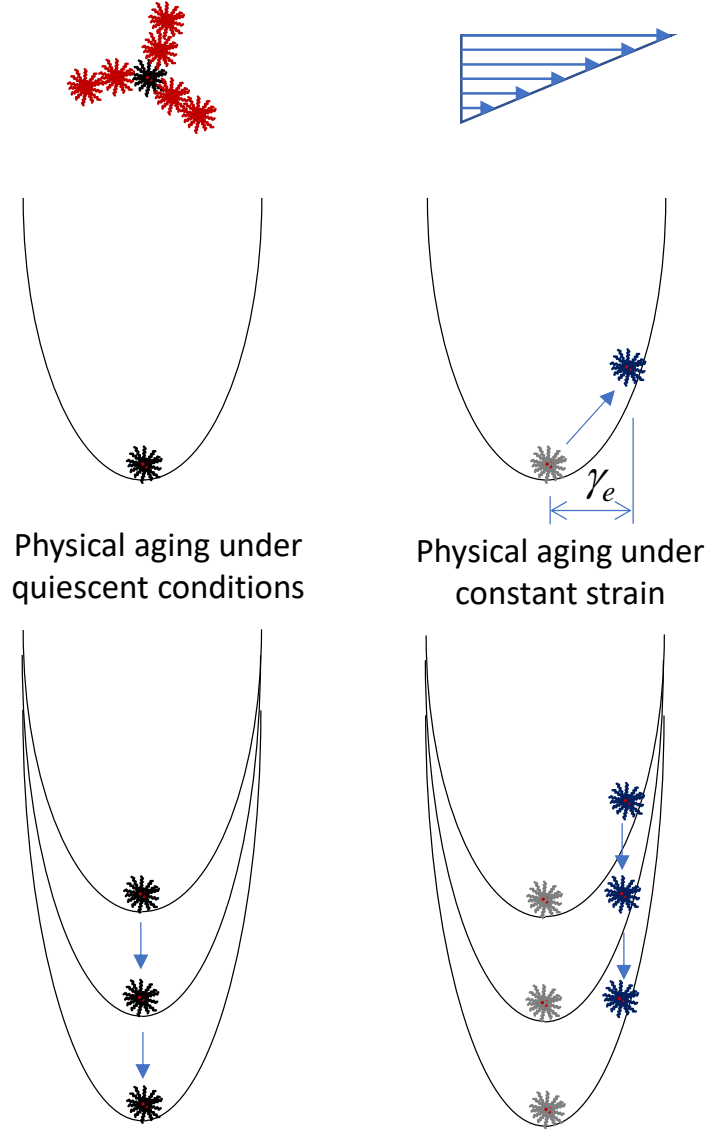

**Figure 12.** Schematic of a structurally arrested particle (a section of a colloidal gel, top left) in an energy well (middle left). In the process of physical aging, the particle explores available phase space and progressively attains those states that take it to the lower well depths as a function of time (bottom left). Under the application of strain

(top right), the particle gains elastic energy (middle right). However, in such a strained state as well, the particle undergoes physical aging wherein it lowers its energy (bottom right).

Eq. (21) suggests that part of the energy that otherwise gets dissipated as the heat gets converted to work under a strained state. This specific case is a limiting case where the dashpot has infinite viscosity. For a material wherein dashpot has finite viscosity, the net work that can be recovered will go on reducing due to viscous dissipation.

In order to analyze how the viscous dissipation causes an increase in free energy, we consider an opposite limit dominated by viscous flow (suggesting physical aging to be negligible under the applied flow field). The rate of viscous dissipation per unit volume under flow is given by $\underline{\underline{\sigma}} : \underline{\underline{\dot{\gamma}}}_v$. However, part of this energy will be used for breaking the structure (rejuvenation or causing an increase in free energy) while the remaining part will dissipate as heat ($\dot{q}_{out,VD}$). Consequently, in a limit of negligible aging and viscous flow the energy balance is given by:

$$\frac{df}{dt} = \underline{\underline{\sigma}}^* : \underline{\underline{\dot{\gamma}}}_v^* - \dot{q}_{out,VD}, \qquad (22)$$

The above energy balance also reiterates that unless breakage of structure occurs, rejuvenation cannot take place. The elastic strain, while can indeed stretch the material and take the system closer to the point of break, only the breakage of structure, which is dissipative in nature, can increase the structure parameter and hence the free energy.

The objective of the present work is to propose a structural kinetic model framework for a thixotropic viscoelastic material, wherein modulus does increase with time without violating the second law of thermodynamics. However, there is a necessity to come up with a comprehensive thermodynamic framework, wherein structure factor $\lambda$ can be explicitly related to the free energy along with its dependence on time, deformation field (rate of dissipation), and mechanical work that can be recovered. A structural kinetic model along with the thermodynamic framework is an open problem for future work. There are many open problems on the experimental front. There are materials that show a continuous increase in relaxation time and elastic modulus as a function of time and are known not to reach equilibrium over the practical timescales.

It is important to understand if such material show non-monotonic stress relaxation, as demonstrated in Fig. 1 but without showing complete stress relaxation. Furthermore, it is also important to understand how the flow history affects the stress relaxation in materials that show physical aging and rejuvenation, particularly with respect to the elastic modulus.

## V. Conclusion

In this work, we address the issue of violation of the second law of thermodynamics for a thixotropic system, wherein elastic modulus increases as a function of time and decreases under an application of the deformation field. We propose a simple structural kinetic model with time-dependent single mode linear Maxwell fluid as a constitutive equation. We show that the second law does not get violated if the rejuvenation is considered to be occurring solely due to the viscous rate of strain in a structural kinetic model. We also show that, for a certain parameter range, leading to time-dependent relaxation time and modulus, along with rejuvenation term depending only on the viscous rate of strain, the model demonstrates non-monotonic stress relaxation as recently reported experimentally. However, the model clearly suggests that the concept of thixo-elasticity is non-existent, and no real material can be termed as thixo-elastic.

**Acknowledgment**: We acknowledge financial support from the Science and Engineering Research Board, Government of India. We thank Prof. Dimitris Vlassopoulos for sharing the experimental data shown in Fig. 1 We thank Prof. Francisco Rubio-Hernandez for constructive comments. We are also indebted to Prof. Ronald Larson for discussion and the critical feedback on various aspects of thixotropy and thermodynamics.